\documentclass[11pt,a4paper]{scrartcl}
\usepackage{url}

\usepackage{multicol}
\usepackage{footmisc}
\usepackage{amsmath}
\usepackage{amssymb}
\usepackage{amsthm}
\usepackage{graphicx}
\usepackage{subfig}
\usepackage{mathabx} % for widebar
\usepackage{dsfont} % for indicator function
\usepackage{multirow} % for table
\graphicspath{{figs/}}

\newtheorem{theorem}{Theorem}

\newcommand{\prob}{\mathbb{P}}
\newcommand{\esp}{\mathbb{E}}

\newcommand{\etal}{\emph{et al.}\ }

\begin{document}
\title{Towards control of opinion diversity by introducing zealots into a polarised social group}

\author{\textbf{Antoine Vendeville} \\ [2ex]
Department of Computer Science\\Centre for Doctoral Training in Cybersecurity\\
University College London, United Kingdom\\\\
\textbf{Benjamin Guedj} \\ [2ex]
Department of Computer Science\\Centre for Artificial Intelligence \\
University College London, United Kingdom\\
Inria, France\\\\
\textbf{Shi Zhou} \\ [2ex]
Department of Computer Science\\Centre for Artificial Intelligence \\
University College London, United Kingdom\\\\
}
\date{}

\maketitle

\begin{abstract}
\textbf{This is the corrected version of a paper that was published as} \textit{A. Vendeville, B. Guedj and S. Zhou, Towards control of opinion diversity by introducing zealots into a polarised social group, Complex Networks \& Their Applications X, pp.\ 341–352, 2022 \cite{vendeville2022}.} \textbf{Since then, a small mistake was found in equation (\ref{backfire_equilibrium}). We fixed the mistake and updated Figures (\ref{optim}) and (\ref{optim_error}). Our overall findings are unchanged.}

We explore a method to influence or even control the diversity of opinions within a polarised social group. We leverage the voter model in which users hold binary opinions and repeatedly update their beliefs based on others they connect with. Stubborn agents who never change their minds (``zealots'') are also disseminated through the network, which is modelled by a connected graph. Building on earlier results, we provide a closed-form expression for the average opinion of the group at equilibrium. This leads us to a strategy to inject zealots into a polarised network in order to shift the average opinion towards any target value. We account for the possible presence of a \emph{backfire effect}, which may lead the group to react negatively and reinforce its level of polarisation in response. Our results are supported by numerical experiments on synthetic data.
\end{abstract}

\section{Introduction}

We are interested in controlling opinions on connected networks with arbitrary degree distribution. In recent years, recommendation algorithms on social platforms have greatly enhanced confirmation bias by showing users content that is the most susceptible to match their interests --- the so-called ``filter bubble'' effect \cite{pariser}. As a consequence more and more isolated, tightly clustered online communities of similar-minded individuals have arisen in various domains such as politics \cite{cota2019,delvicario2017,garimella2018}, healthcare \cite{allington2020,health_polarisation} or science \cite{williams2015}. Because of the so-called \emph{backfire effect}, presenting these users with opposing information might have the adverse effect of reinforcing their prior beliefs \cite{bail2018,schaewitz2020}. In this paper we provide a simple method to shift diversity of opinions within towards a chosen target level, with and without the presence of a backfire effect. 

To this end we rely on the well-known voter model, in which each user holds one of two possible opinions (e.g.\ liberal of conservative, pro or anti-abortion) and updates it randomly under the distribution of others' beliefs. Independently introduced by Clifford and Sudbury \cite{clifford_sudbury} and Holley \cite{holley1975} in the context of particles interaction, this model has since been used to describe in a simple and intuitive manner social dynamics where people are divided between two parties and form their opinion by observing that of others around them. We assume some of the users are stubborn and never change opinion. We call them \emph{zealots} as in \cite{mobilia2003,mobilia2007}. They can represent lobbyists, politicians or activists for example. Long time dynamics and limiting behaviour of such processes have been subject to several studies \cite{mobilia2007,mukhopadhyay2020,binary_opinion}. 

To achieve our goal we extend a previous result from the literature. Namely, authors of \cite{mobilia2007} found an expression for the average number of opinion-1 users at equilibrium in the $n\rightarrow\infty$ limit for a fully-connected network. Here we prove that their result also holds on expectation for any connected graph, assuming the placement of zealots is done uniformly at random. This allows us to find the optimal number of zealots to inject in a polarised community in order to reach any target average opinion at equilibrium, with and without the presence of a backfire effect. This effect we model simply by assuming that the injection of any number of zealots entails the \emph{radicalisation} of some non-zealous users, turning them into zealots with the opposite opinion.

Our findings are illustrated through numerical simulations. Four different user graph topologies are considered: a fully-connected group where everyone is influenced by everyone else, an Erdös-Rényi random graph, a Barabási-Albert scale-free graph and a Watts-Strogatz small-world network. We find empirical averages to be close approximations of theoretical values. All code used for the simulations is available online.\footnote{\url{https://github.com/antoinevendeville/howopinionscrystallise}}

\section{Related Literature}
Perhaps the earliest milestone in the study of opinion dynamics are the works from French \cite{french} and Degroot \cite{degroot} who studied how a society of individuals may or may not come to an agreement on some given topic. Assuming the society is connected and people repeatedly update their belief by taking weighted averages of those of their neighbours, they showed that consensus is reached. That is, everyone eventually agrees. Various other models have been developed since, to tackle the question of under which circumstances and how fast a population is able to reach consensus. Amongst others, \cite{friedkin_johnsen} introduces immutable innate preferences, \cite{axelrod} studies the effect of homophily, \cite{word_of_mouth} assumes individuals are perfectly rational and \cite{nbsl} accounts for the influence of external events. 

The voter model was introduced independently by Clifford and Sudbury \cite{clifford_sudbury} and Holley \cite{holley1975} in the context of particles interaction. They proved that consensus is reached on the infinite $\mathbb{Z}^d$ lattice. Several works have since looked at different network topologies, wondering whether consensus is reached, on which opinion and at what speed. Complete graphs \cite{yehuda2002,perron2009,sood2008,yildiz2010}, Erdös-Rényi random graphs \cite{sood2008,yildiz2010}, scale-free random graphs \cite{fernley2019,sood2008}, and other various structures \cite{sood2008,yildiz2010} have been addressed. Variants where nodes deterministically update to the most common opinion amongst their neighbours have also been studied \cite{chen2005,mossel2013}.

An interesting case to consider is the one where zealots -- i.e.\ stubborn agents who always keep the same opinion, are present in the graph. Such agents may for example represent lobbyists, politicians or activists, i.e.\ entities looking to lead rather than follow and who will not easily change side. One of those placed within the network can singlehandedly change the outcome of the process \cite{mobilia2003,sood2008}. If several of them are present on both sides, consensus is usually not reachable and instead opinions converge to a steady-state in which they fluctuate indefinitely \cite{mobilia2007,binary_opinion}.

Recently, Mukhopadhyay  \cite{mukhopadhyay2020} considered zealots with different degrees of zealotry and proved that time to reach consensus grows linearly with their number. They also showed that if one opinion is initially preferred --- i.e.\ agents holding that opinion have a lesser probability of changing their mind --- consensus is reached on the preferred opinion with a probability that converges to 1 as the network size increases. Klamser \etal \cite{klamser2017} studied the impact of zealots on a dynamically evolving graph, and showed that the two main factors shaping their influence are their degrees and the dynamical rewiring probabilities. 

With the increasing importance of social networks in the political debate and information diffusion, there has been a recent surge in research aiming at controlling opinions, often with the goal to reduce polarisation. In the context of the Voter and Friedkin-Johnsen models respectively, Yildiz \etal \cite{binary_opinion} and Goyal \etal \cite{goyal2019} provide algorithms for selecting an optimal sets of stubborn nodes in order to push opinions in a chosen direction. Our work differs from the former in that we propose a more general strategy, that works on expectation for any connected graph. Yi and Patterson \cite{yi2019disagreement} formulate different constrained optimisation problems under the French-Degroot and the Friedkin-Johnsen models. They provide solutions in the form of optimal graph construction methods. 

Still within the Friedkin-Johnsen paradigm, Chitra and Musco \cite{chitra2020} prove that dynamically nudging edge weights in the user graph can reduce polarisation while preserving relevance of the content shown by the recommendation algorithm. Garimella \etal \cite{garimella2017bis} propose a method to reduce polarisation through addition of edges in the network. The focus is put on which nodes to connect in order to get the best reduction in polarisation, while being sure that the edge is ``accepted'' --- as extreme recommendations might not work because of the backfire effect. Finally, Cen and Shah \cite{cen2020} propose a data-driven procedure to moderate the gap between opinions influenced by a neutral or a personalised newsfeed. Importantly, they show that this can be done even without knowledge of the process through which opinions are derived from the newsfeed.

\section{The Voter Model with Zealots} \label{voter_model}

Consider a graph with $n$ users labelled $1, \ldots, n$ that can each hold opinion 0 or 1. Given an initial distribution of opinions, each user updates their opinion at the times of an independent Poisson process of parameter 1 by adopting the opinion of one of its neighbours chosen uniformly at random. Letting $x_i(t)$ denote the opinion of user $i$ at time $t$, we say that consensus is reached if almost surely all users eventually agree, i.e.\ if
\begin{equation} \label{consensus}
	\forall i,j, \quad \prob \left(x_i(t) = x_j(t) \right) \underset{t \rightarrow \infty}{\longrightarrow} 1.
\end{equation}
On any finite connected network, consensus is reached \cite{aldous_fill_2014}. Intuitively, no matter the current number of opinion-0 and opinion-1 users, there exists a succession of individual opinion changes with strictly positive probability that results in everyone holding the same opinion.

Here we place ourselves in the case where the user network is a connected graph with arbitrary degree distribution. This means there is a path from any user to another, and degrees of different users are not correlated with one another. Edges may be directed or not. We are interested in the particular situation where some of the agents are stubborn and never change their opinions. We call such agents \emph{zealots}. Whenever a clock associated with a zealot rings, their opinion is not updated. They form an inflexible core of partisans within a group who bear great power of persuasion over the whole population. Both the position of these agents and initial opinions are assumed to be independent from the network topology.

\begin{figure}
    \centering
    \subfloat{\includegraphics[width=.45\textwidth]{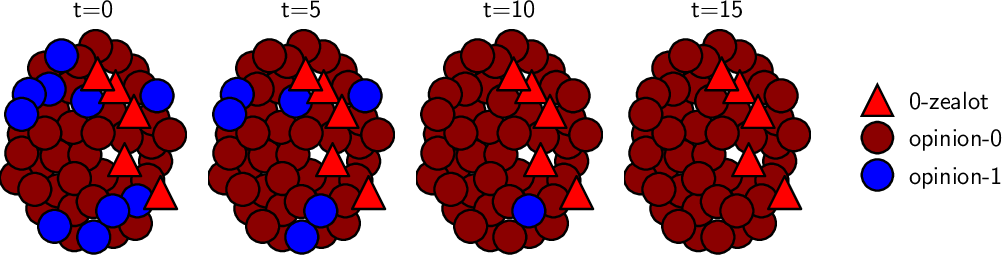}}~
    \subfloat{\includegraphics[width=.45\textwidth]{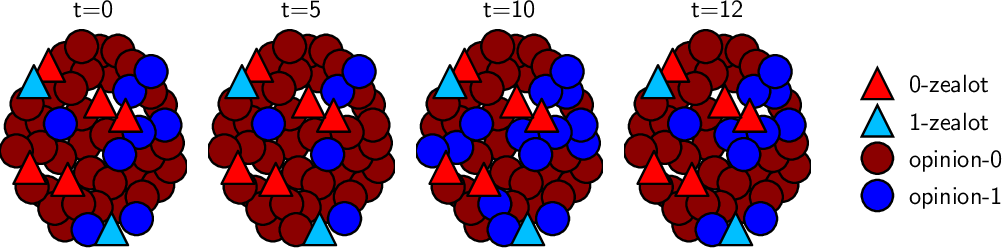}}
    \caption{\textbf{(Left)} Example realisation of the model on a complete graph at different times with $n=50,n_1=10,z_0=5,z_1=0$. Because there are no 1-zealots, everyone eventually adopts opinion 0. \textbf{(Right)} Same setting except $z_1=2$. Because there are zealots in both camps, the system reaches a state of equilibrium where no opinion prevails.}
	\label{voter_graph}
\end{figure}

We call 1-zealot a stubborn agent with opinion 1 and denote by $z_1$ their number. Similarly, $z_0$ will denote the quantity of 0-zealots. The remaining $n-z_0-z_1$ users are free to change opinions during the whole duration of the process. If $z_0>0,z_1=0$ or $z_0=0,z_1>0$ then via similar arguments as for eq.~\ref{consensus} consensus is reached on opinions 0 and 1 respectively. Here we are particularly interested in the case $z_0,z_1>0$, where there is always a strictly positive number of users with each opinion. This prevents consensus and instead the system reaches state of equilibrium in which it fluctuates indefinitely \cite{mobilia2007,binary_opinion}. We illustrate cases $\{z_0>0,z_1=0\}$ and $\{z_0,z_1>0\}$ in fig.~\ref{voter_graph}. 

Our results are valid as long as at least one of $z_0$ and $z_1$ is strictly positive and we formally require $z_0+z_1>0$. Importantly, we assume that for any fixed tuple $(z_0,z_1)$, the position of zealots is drawn uniformly at random. Letting $Z$ be a random vector encoding these positions we formally write $Z\sim\mathcal{U}$.

%We write $[m_{ij}]_{i,j}$ to denote the matrix with entry $m_{ij}$ in the $i$-th row and $j$-th column and let $e^M$ denote the exponential of any matrix $M$. Finally if $\Omega$ is a set we use $\mathcal{P}(\Omega)$ to indicate its powerset.

\section{Expected Opinion Diversity at Equilibrium} \label{distrib_and_equilibrium}
Assuming $z_0+z_1>0$, $N_1(t)$ converges to a state of equilibrium, which is characterised by a time-independent stationary distribution $\pi$. If $z_0=0$ (resp.\ $z_1=0$) then $\pi$ is the constant distribution $\delta_n$ (resp.\ $\delta_0$). This equilibrium does not depend on the initial opinions of non-zealots but on the topology of the graph and the position of zealots. Let $N_1^\star$ be a random variable distributed under $\pi$. Its average value informs us on the limiting opinion diversity of the group. Authors of \cite{mobilia2007} proved that this average was $nz_1/(z_0+z_1)$ for complete graphs in the $n\rightarrow \infty$ limit. We now show that this results holds on expectation for any connected graph where the position of zealots is drawn uniformly at random.
\begin{theorem} \label{limit_theo}
For any connected user graph and any $z_0,z_1$ such that $z_0+z_1>0$, we have
\begin{equation} \label{limit_formula}
	\esp_{Z\sim\mathcal{U}} \esp \left[N_1^\star\right] = n\frac{z_1}{z_0+z_1}.
\end{equation}
%In the special case where the graph is complete, all possible positionings of zealots are equivalent so that we have simply
%\begin{equation} \label{limit_formula_complete}
%	\esp \left[N_1^\star\right] = n\frac{z_1}{z_0+z_1}.
%\end{equation}
\end{theorem}
This theorem states that the proportion of opinion-1 users is expected to endlessly fluctuate around the ratio $z_1/(z_0+z_1)$. For example, having twice as many zealots as the other camp will on average lead to count twice as many partisans. Thus the camp that boasts the biggest quantity of zealots is expected to be of bigger size in the long run. The ratio $z_1/(z_0+z_1)$ does not depend on $n$ and can be interpreted as the average opinion diversity at equilibrium. Finally, if the graph is complete then all possible values of $Z$ are equivalent and the left side of  eq.~\ref{limit_formula} is simply $\esp \left[N_1^\star\right]$. %Finally, it should not come as surprising that if $z_0=0$ (resp.\ $z_1=0$), the system converges to the static state where everyone holds opinion 1 (resp.\ opinion 0).

\begin{proof}[Theorem~\ref{limit_theo}]
	From \cite[Theorem 2.1]{binary_opinion}, the vector of individual opinions $(x_1(t), \ldots, x_n(t))$ converges in distribution to a random vector $(x_1^\star, \ldots, x_n^\star)$. From \cite[Proposition 3.2]{binary_opinion} we have that $\esp[x_i^\star]$ is equal to the probability that a random walk on the user graph initiated at node $i$ is absorbed by the set of 1-zealots. Because here we consider a connected graph with $z_0,z_1$ zealots placed uniformly at random, it holds that $\esp_{Z\sim\mathcal{U}}\esp[x_i^\star] = z_1/(z_0+z_1)$. There are $n-z_0-z_1$ non-zealots and $z_1$ 1-zealots, thus we have
\begin{equation}
	\esp_{Z\sim\mathcal{U}}\esp \left[N_1^\star\right] = z_1 + (n-z_0-z_1) \esp_{Z\sim\mathcal{U}}\esp[x_i^\star]
\end{equation}
and eq.~(\ref{limit_formula}) ensues.
\end{proof}

\subsection{Empirical Approximation}
Let briefly discuss how to verify Theorem~\ref{limit_theo} through computer simulations. Assume we perform $M$ simulations of our model on a given graph. Each time $Z$ is drawn at random under $\mathcal{U}$ at the beginning. Then the law of large numbers tell us that
\begin{equation} \label{simu_approx}
	\frac{1}{M} \sum_{m=1}^M \esp \left[N_1^\star\right] \underset{M\rightarrow\infty}{\longrightarrow} n\frac{z_1}{z_0+z_1}. 
\end{equation}
Now $N_1^\star$ is a random variable distributed under $\pi$. For a given realisation of $N_1(t)$, its expectation can be approximated empirically via 
\begin{equation} \label{time_approx}
	\frac{1}{T} \sum_{t \in \mathcal{T}} N_1(t) \underset{T\rightarrow\infty}{\longrightarrow} \esp \left[N_1^\star\right]
\end{equation}
where $\mathcal{T}$ is a set of sufficiently large times and $T$ denotes its size. Combining (\ref{simu_approx}) and (\ref{time_approx}) we have:
\begin{equation}
	\frac{1}{MT} \sum_{m=1}^M \sum_{t \in \mathcal{T}} N_1(t) \underset{M,T\rightarrow\infty}{\longrightarrow} n\frac{z_1}{z_0+z_1}.\end{equation}
Furthermore, we can also approximate the expected stationary distribution:
\begin{equation} \label{proba_simu}
	\frac{1}{MT} \sum_{m=1}^M \sum_{t \in \mathcal{T}} \mathds{1}\{N_1(t)=k\} \underset{M,T\rightarrow\infty}{\longrightarrow} \esp_{Z\sim\mathcal{U}} [\pi_k].
\end{equation}

\section{Control of Opinion Diversity}
Theorem~\ref{limit_theo} is a useful tool for the control of opinions diversity at equilibrium. Let us represent a polarised online community by a connected graph of $n$ individuals with a quantity $z_0>0$ of 0-zealots and no 1-zealot ($z_1=0$). For the sake of clarity we assume the graph to be complete but the strategy holds on expectation when $Z\sim\mathcal{U}$ in the more general case of any connected graph. The completeness assumption is also fairly reasonable for certain social platforms, such as Facebook or Reddit where everyone in a group or subreddit sees the posts of everyone else. 

Because $z_0>0$ and $z_1=0$, the community is homogeneous: each member will end up adopting opinion 0 no matter what. The presence of such polarised groups hinder democratic debate online, as they prevent the evolution of opinion and reinforce pre-existing beliefs. To mitigate this phenomenon, we suggest injecting 1-zealots into the group. Doing so means forcing $z_1>0$ and thus at equilibrium, the group will be more diverse and divided between the two camps. 

From Theorem~\ref{limit_theo}, the average opinion at equilibrium is given by $z_1/(z_0+z_1) \in [0,1]$ and equals 0 here. Let us choose a target value $0<\lambda<1$, representing the level of diversity we want the group to reach. Typically it should be around $1/2$ if the goal is to transform the echo chamber into a diverse sphere of opinions. It is immediate that injecting a quantity
\begin{equation} \label{s1_optimal}
	z_1^\star = \frac{\lambda}{1-\lambda} z_0
\end{equation}
of 1-zealots yields $z_1^\star/(z_0+z_1^\star) = \lambda$. In practice we round this value to the previous or next integer. If $\lambda=1/2$, then we should add exactly as many 1-zealots as there are 0-zealots. 

Note that the equilibrium opinion is only determined by the quantity of zealots on each side and other users do not affect it. The expected opinion held by such a user converges to $z_1/(z_0+z_1)$ and thus their number and initial opinions does not impact the equilibrium. Hence, one could ``convert'' users amongst the initial $n-z_0$ non-zealots into 1-zealots instead of adding external ones. That would lead to the same diversity, except that it comes with a limit on the quantity of 1-zealots because of the constraint $z_0 + z_1 \leq n$. Thus in that case, our strategy is only feasible if
\begin{equation} 
	z_0 + \frac{\lambda}{1-\lambda} z_0\leq n.
\end{equation}
In a more general fashion, if we are limited in the number of zealots we can add to the network by $z_1^{\text{max}}$, the optimal quantity of such agents is given by:
\begin{align} 
	z_1^\star = \text{min} \left(z_1^{\text{max}}, \frac{\lambda}{1-\lambda} z_0 \right).
\end{align}
%Additionally, time constraints may be given in the form of a maximum duration tolerated for the group to reach the target diversity. Because convergence time of the model decreases with the number of zealots (cv.\ Theorem~\ref{cvtime_theo}), this comes down to imposing a minimum number of 1-zealots to be added into the network.

\subsection{Backfire Effect} Numerous studies suggest that presenting certain users with opposing views might actually entrench them even deeper in their beliefs. This is known as the backfire effect. To account for it we study the scenario where in reaction to 1-zealots being created (either by addition or conversion), some of the non-zealous users will \emph{radicalise} and become 0-zealots. Formally, we set that for each increment of $z_1$, a quantity $\alpha<1$ of non-zealous users become 0-zealots. The opinion at equilibrium is now given by
\begin{equation} \label{backfire_equilibrium}
	\frac{z_1}{z_0+z_1+\alpha z_1}
\end{equation}
and diversity $\lambda$ is exactly reached with
\begin{equation} \label{s1star_backfire}
	z_1^\star = \lambda z_0D^{-1}
\end{equation}
where $D:=1-\lambda -\lambda \alpha$. If $D>0$ then $z_1^\star>0$ and we can inject this quantity of users into the network. If $D\leq0$ however this becomes impossible as (\ref{s1star_backfire}) is then either undefined or negative. In this case, we find that 
\begin{equation}
	z_1 \mapsto \left(\frac{z_1}{z_0+z_1+\alpha z_1} - \lambda \right)^2
\end{equation}
is strictly positive and decreasing towards $((1+\alpha)^{-1}-\lambda)^2$ over $\mathbb{R}_{>0}$. Thus the larger $z_1$ the closer we get to the target diversity, up to a certain point. This means that the addition of 1-zealots entails the radicalisation of too many users into 0-zealots for it to ever counterbalance the spread of opinion 0. If we are limited in the number of zealots we can add to the network by $z_1^{\text{max}}$, we have the natural optimal values for $z_1$:
\begin{equation} \label{optim_backfire}
	\begin{cases}
		z_1^\star = \text{min} \left(z_1^{\text{max}}, \lambda z_0D^{-1} \right) &\text{ if } D>0,\\
		z_1^\star = z_1^{\text{max}} &\text{ if } D\leq0.
	\end{cases}
\end{equation}
Note that in the case where we are converting existing users into zealots instead of injecting external ones, $z_1^{\text{max}}$ is given by the constraint
\begin{equation}
	z_1^{\text{max}} + \underbrace{z_0+\alpha z_1^{\text{max}}}_{\substack{\text{updated $z_0$} \\ \text{after backfire}}} = n. \label{z1max}
\end{equation}

\section{Numerical Experiments} \label{simu_section}
\begin{figure} 
	\centering
	\includegraphics[width=.92\textwidth]{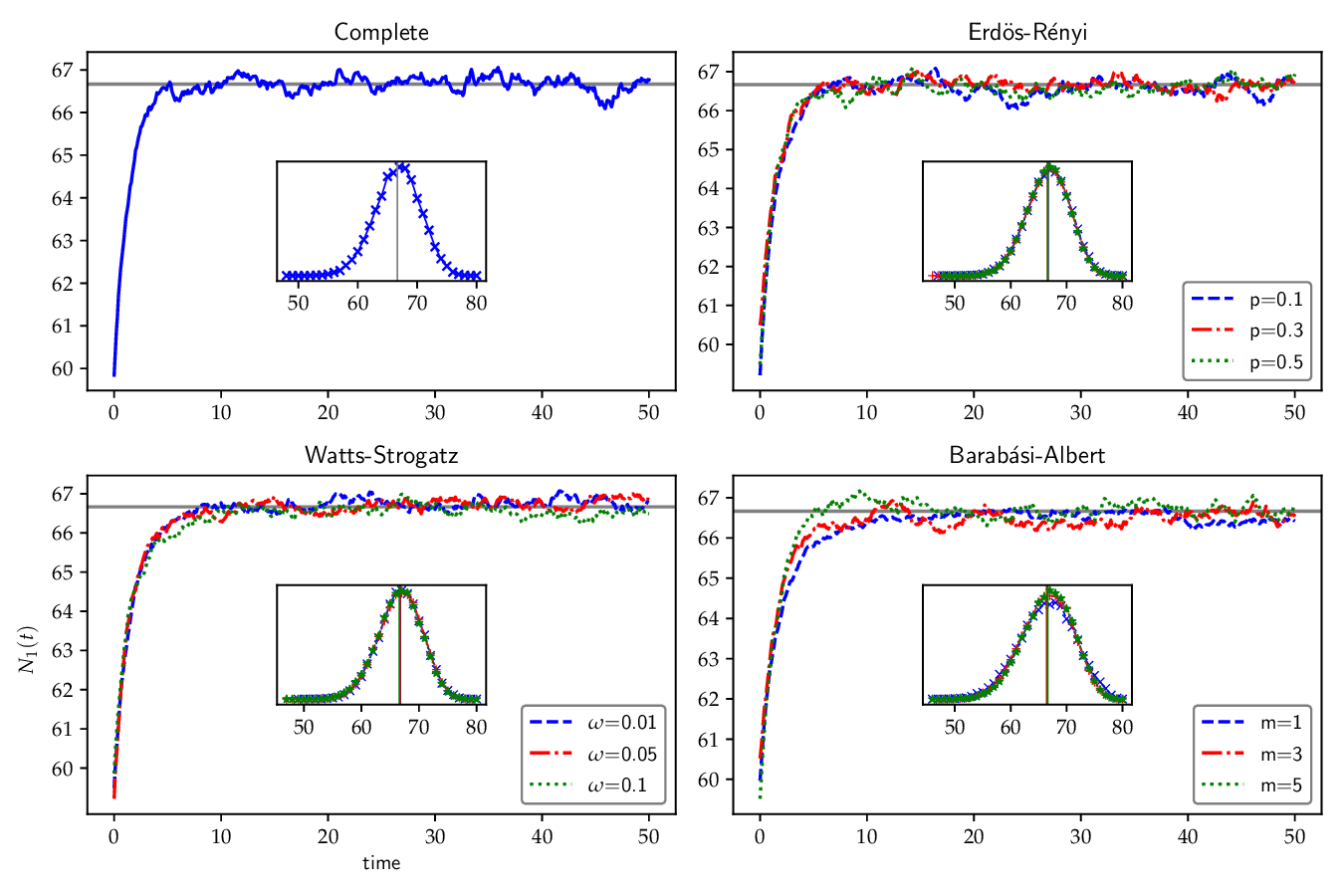}
	\caption{Evolution of $N_1(t)$ for various graph models and parameters. Values are averaged over 500 simulations. Horizontal grey lines indicate the limiting expectation given by Theorem~\ref{limit_theo}. Insets show empirical distributions at equilibrium.}
	\label{N1t_evolve}
\end{figure}

We now validate Theorem~\ref{limit_theo} via computer simulations. Let us set $(n,z_0,z_1)=(100,20,40)$ and $\phi=nz_1(z_0+z_1)\simeq66.7$. We consider four different undirected graph models with varying parameters:
\begin{itemize}
	\item Complete graph with an edge between each pair of users, 
	\item Erdös-Rényi random graph with density $p=0.1, 0.3, 0.5$,
	\item Watts-Strogatz small-world graph with initial connections to the 4 nearest neighbours and rewiring probability $\omega=0.01,0.05,0.1$,
	\item Barabási-Albert scale-free graph with $m=1,3,5$ initial nodes and $m$ new connections at each step.
\end{itemize}
For each model and each parameter we generate one connected user graph then perform $M=500$ simulations. At the beggining of each, position of zealots and initial opinions are drawn uniformly at random, then we let the model evolve for 200 time units. In Fig.~\ref{N1t_evolve} we plot values of $N_1(t)$ averaged over all simulations for each graph model and parameter---up to $t=50$ and not $200$ for the sake of clarity. We observe a good correspondence with the theory, as after about $20$ time units all the empirical averages seem to fluctuate closely around $\phi$. %Values of $N_1(t)$ are saved every $0.1$ time unit.

Insets show empirical distributions of $N_1^\star$ obtained via eq.~\ref{proba_simu}. For any given simulation, the empirical distribution is obtained by averaging all values of $N_1(t)$ starting at $t=20$. This value was chosen by manual inspection, to make sure that the system had enough time to stabilise. Distributions for each choice of graph model and parameter are then averaged over all 500 simulations. We also show their means and its theoretical value $\phi$. We observe very good correspondence between all the Erdös-Rényi and Watts-Strogatz graphs for all parameters, and a slightly worse on for the Barabási-Albert graphs. This might be due to the lower regularity in the graph topology, as when $m$ increases, the distribution seem to get closer to the others. Finally, the means are all close to the theoretical value.

%\begin{figure} 
%	\centering
%	\includegraphics[width=\textwidth]{multigraphs_distrib_equilibrium.eps}
%	\caption{Empirical distribution of $N_1^\star$, averaged over 500 simulations after 20 time units. Colored vertical bars represent empirical means and grey vertical bars represent the unique theoretical mean $\phi$.}
%	\label{plot_distrib}
%\end{figure}

To quantify more precisely the discrepancy between theory and simulations, we look more closely at the difference between empirical $N_1(t)$ and $\phi$. Again we only consider results from $t=20$ onwards. For any graph model and parameter, the average relative error of the empirical means with respect to $\phi$ is calculated through:
\begin{equation}
	\frac{1}{\phi} \frac{1}{M} \sum_{m=1}^M \left\vert \frac{1}{T} \sum_{t=20}^{100} N_1(t) -\phi \right\vert
\end{equation}
where $M=500$ is the number of simulations and $T$ is the number of data points from $t=20$ onwards. This value quantifies the average gap between empirical means of $N_1(t)$ at equilibrium and their theoretical value $\phi$, expressed as a percentage relative to the value of $\phi$. Results are presented in Table~\ref{error_std_table} alongside standard deviations. %\footnote{Because we saved empirical values of $N_1(t)$ every 0.1 time unit, we have $T=10\times(100-20+1)$.}

As to be expected the complete network boasts the most precise results, with an average relative error of $0.8\%\pm0.6\%$ corresponding to a difference of $0.5\pm0.4$ users. Then in order of decreasing accuracy we have successively Erdös-Rényi, Watts-Strogatz and finally Barabási-Albert graphs. Each time the precision increases with the density of the graph, as the network gets closer and closer to the complete case. The worst case is the Barabási-Albert network with parameter $m=1$, yielding an average relative error of $4.7\%\pm3.5\%$ which corresponds to a difference of $3.1\pm2.3$ users.

\begin{table}
    \setlength{\tabcolsep}{15pt}
    \centering
    \caption{Relative errors between empirical averages of $N_1(t)$ at equilibrium and limiting expectation $nz_1/(z_0+z_1)$, averaged over 500 simulations and with standard deviation.}
    \begin{tabular}{lcccl}
    \cline{1-4}
    \textbf{Complete} & $0.8\%\pm0.6\%$ & - & - &  \\ \cline{1-4}
    \multirow{2}{*}{\textbf{Erdös-Rényi}} & $p=0.1$ & $p=0.3$ & $p=0.5$ &  \\
     & $1.7\%\pm1.2\%$ & $1.0\%\pm0.7\%$ & $0.9\%\pm0.6\%$ &  \\ \cline{1-4}
    \multirow{2}{*}{\textbf{Watts-Strogatz}} & $\omega=0.01$ & $\omega=0.05$ & $\omega=0.1$ &  \\
     & $2.5\%\pm1.9\%$ & $2.6\%\pm1.9\%$ & $2.4\%\pm1.9\%$ &  \\ \cline{1-4}
    \multirow{2}{*}{\textbf{Barabási-Albert}} & $m=1$ & $m=3$ & $m=5$ &  \\
     & $4.7\%\pm3.5\%$ & $3.3\%\pm2.4\%$ & $2.8\%\pm2.1\%$ &  \\ \cline{1-4}
    \end{tabular}
    \label{error_std_table}
\end{table}

%We also look at the unbiased relative standard deviation, given by
%\begin{equation}
%	\sqrt{\frac{1}{M-1} \sum_{m=1}^M \left( \frac{ \frac{1}{T}\sum_{t=20}^{100} N_1(t) -\frac{nz_1}{z_0+z_1}}{nz_1/(z_0+z_1)} \right)^2}
%\end{equation}

\begin{figure}[t]
	\centering
	\includegraphics[width=\textwidth]{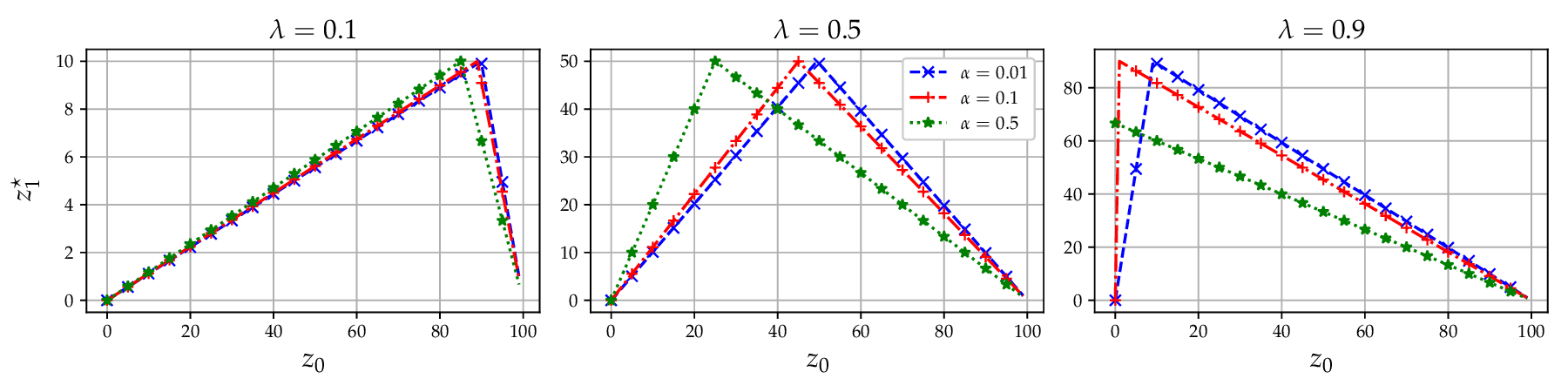}
	\caption{Controlling diversity of opinions with backfire effect. Optimal $z_1^\star$ function of $z_0$ for $n=100$, $\lambda=0.1,0.5,0.9$ and various intensities of the backfire effect $\alpha$. Note that axes scales differ from one plot to another.}
	\label{optim}
\end{figure}

\begin{figure}[t]
	\centering
	\includegraphics[width=\textwidth]{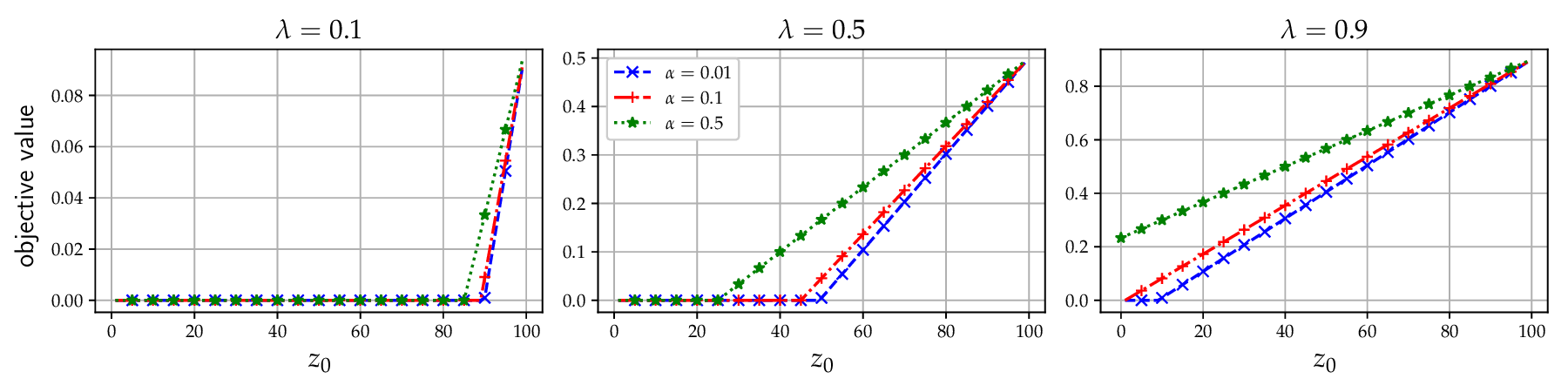}
	\caption{Absolute difference between diversity reached with $z_1^\star=z_1^\text{max}$ and target $\lambda$. Note that axes scales differ from one plot to another.}
	\label{optim_error}
\end{figure}

Finally, we turn to the problem of controlling the diversity of opinions under the presence of a backfire effect. We consider a complete graph of $n=100$ users with $z_0>0, z_1=0$ and three different target diversities $\lambda=0.1, 0.5, 0.9$. The intensity of the backfire effect takes various values  $\alpha \in \{0.01, 0.1, 0.5\}$. The optimal quantity of 1-zealots is calculated according to (\ref{optim_backfire}), in the case where we are converting existing users and thus $z_1^\text{max}$ is given by eq.~\ref{z1max}. In fig.~\ref{optim} we plot $z_1^\star$ function of $z_0$ for each $\alpha$. Additionally in fig.~\ref{optim_error} we plot the error, that is the absolute difference between the diversity reached using $z_1^\star$ and the target $\lambda$.

As expected, for all $\lambda$ considered, the lower $\alpha$ the higher quantities of 0-zealots it is possible to fight against. Peaks in the curves of fig.~\ref{optim} correspond to tipping points after which $z_1^\star=z_1^\text{max}$, either because of the size constraint (\ref{z1max}) or because $D$ becomes non-positive. Thus diversity is reached exactly only before those peaks, as can be seen with nil errors in fig.~\ref{optim_error}. Notice that this peak does not necessarily appear---for example with $(\lambda,\alpha)=(0.9,0.5)$, target diversity is never reached exactly. Finally, note that if we are looking to maximise the diversity of opinions---\emph{i.e.}\ $\lambda=1/2$, errors are rather small for a large range of $z0$ values, meaning it is possible even in the presence of a backfire effect.

\section*{Conclusion and Future Work} \label{futurework}
In this paper we analysed the voter model with zealots on connected graphs with arbitrary degree distribution. Assuming that initial opinions and position fo zealots are drawn uniformly at random, we extended existing results to provide closed-form expressions from \emph{(i)} expected opinion distribution at equilibrium, and \emph{(ii)} expected convergence time. We then used our findings to propose a simple method for the control of opinion diversity in a connected group of users that may or may not be subject to a backfire effect. Our analysis was supported through numerical simulations. Leads for further work include extensions such as considering more than two opinions or implementing variable degrees of zealotry.

%
% ---- Bibliography ----
%
\bibliographystyle{unsrt}%\biboptions{authoryear}
\bibliography{biblio}

\end{document}